# Superconductivity in ternary pyrite-type compound IrBi$_{1-x}$Te$_{1+x}$ ($x \approx 0.2$) at ambient and high pressure

Qing-Ge Mu, Walter Schnelle, Guo-Wei Li, Horst Bormann,

Claudia Felser, Sergey Medvedev[*]

[1]Max Planck Institute for Chemical Physics of Solids, 01187 Dresden, Germany

*Corresponding author, E-mail: sergiy.medvediev@cpfs.mpg.de

**Abstract**

We report superconductivity in the ternary compound IrBi$_{0.8}$Te$_{1.2}$ with critical temperature 1.7 K. The replacement of Bi by Te leads to the metallic conductivity in contrast to the semiconducting parent compound IrBiTe. The superconductivity can be further enhanced by application of external pressure. $T_c$ demonstrates dome-shaped dependence on pressure with a maximum value of 2.6 K at 26.5 GPa. Pressure effects on electronic properties and lattice dynamic of IrBi$_{0.8}$Te$_{1.2}$ were studied by pressure dependent electrical resistivity, Hall effect and Raman spectroscopy measurements.

**Introduction**

The compounds of $TX_2$ family (where $T$ is transition metal, $X$ is chalcogen or pnictogen) with pyrite-type structure (named after mineral pyrite $FeS_2$) have attracted considerable attention as systems close to insulator-metal transition, with various magnetic and electrical phenomena ranging from the antiferromagnetic insulating materials to the superconductors [1-3]. Recently, topological properties are explored in this system [4-8]. $PtBi_2$ was predicted as a Dirac semimetal, and superconductivity is uncovered under pressure [5]. Angel-resolved photoemission spectroscopy studies reveal that $CoS_2$ is a true half-metal and hosts Weyl fermions [8].

The emerging superconductivity with enhanced critical temperature $T_c$ is strongly correlated with various kinds of ordered electronic states or structural instability. The instability of $X$-$X$ anion dimer in pyrite-type structure is proposed to play a significant role in tuning the superconductivity. The highest value of $T_c$ in $Ir_xX_2$ and $Ir_{1-x}Rh_xSe_2$ is achieved when $X$-$X$ separation in the anion dimer is the largest [9,10]. For $PdSe_2$, pressure induces a superconducting transition accompanied with a structural transition from $PdS_2$-type to pyrite-type crystal structure [4]. Moreover, the superconducting $T_c$ shows dome-shaped pressure dependence and reaches maximum at the pressure where anion bond softening is observed [4]. Similar to $PdSe_2$, superconductivity is observed in high-pressure pyrite phase of $PdS_2$, while no phonon softening is detected related to the dome-shaped pressure dependence of $T_c$ [11]. Furthermore, studies on $NiSe_2$ under pressure reveal the same instability of Se-Se dimer under pressure without superconductivity [11].

Superconducting properties in relation to bonding and structural properties of structurally related Ir-Te compounds ($IrTe_2$, $Ir_3Te_8$, $Ir_xTe_2$) have been investigated recently [9,10,12-16]. $IrTe_2$ belongs to two dimensional $CdI_2$-type structure at room temperature, and evolves into monoclinic structure at ~ 250 K [12]. Intercalation of Pd or Cu and chemical doping with Pd or Pt suppress the structural transition and induce superconductivity, which is supposed to be due to bond breaking [12-14]. $Ir_3Te_8$ single crystal undergoes a structural transition from pyrite-type structure to rhombohedral structure at 350 K, and bulk superconductivity is observed at 1.8 K [15]. However, all these compounds undergo a major structural transition instead of progressive variation of the anion dumbbell. Another degree of freedom in crystal chemistry and tuning the properties is offered in the ternary compounds $TYX$ (where $T$ is transition metal, $Y$ is pnictogen and $X$ is chalcogen) most of which (over a hundred are known) crystallize in the pyrite (or closely related) structures containing bond anion pairs $X_2$, $Y_2$ or $XY$ [17,18]. Such ternary compounds of Co-group metals are of interest since their number of valence electrons is the same as that of the pyrite $FeS_2$. Therefore, the electronic configuration in these ternary compounds should be the same as pyrite $FeS_2$, and like the pyrite they should be semiconducting. Indeed, all the obtained Rh and Ir compounds in this family are semiconductors [17,18]. The possibility to tune the electronic or crystalline structure in a wide range by chemical substitution or applying external pressure make this family of compounds a suitable playground for investigating the interplay between crystal structure and electronic properties, which provides further understanding of the underlaying mechanism of superconductivity in pyrite-type materials.

Here we report the synthesis of ternary $IrBi_{1-x}Te_{1+x}$ ($x \approx 0.2$) crystals and studies of its electronic transport properties at ambient pressure and high pressure. The compound appears to be metallic and superconducting with $T_c$ 1.7 K at ambient pressure. The $T_c$ can be further increased by the applied high pressure delivering a dome-shaped pressure dependence of $T_c$.

**Experimental details**

The single crystals of named $IrBi_{0.5}Te_{1.5}$ were grown out of Bi flux. The mixture of high purity Ir powder (99.9%), Bi powder (99.999%), and Te powder (99.9999%) with atomic ratio Ir:Bi:Te = 2:4:3 were weighed accurately and sealed in an evacuated quartz tube ($10^{-4}$ Torr). The sealed quartz tube was kept vertically inside an automated programmable box furnace. It was heated to 900 °C within 24 h, held for 24 h, and then slowly cooled (1 °C h$^{-1}$) down to 600 °C with a hold time of 24 h. Further, the sample was cooled from 600 °C to room temperature naturally. Cubic crystals up to hundreds of micrometers were obtained by removing the extra Bi flux with centrifuging at 500 °C. The as grown crystals were immersed in nitric acid ($\geq$ 65%) for 3 h to remove the residual Bi flux. The chemical composition was analyzed by energy dispersive x-ray spectroscopy (EDX), and the crystal structure was characterized by single crystal x-ray diffraction (SXRD) at room temperature.

The electrical resistivity and heat capacity measurements at ambient pressure were carried out on PPMS-9 equipped with He$^3$ cooling system with temperature down to 0.34 K. For heat capacity measurements, several pieces of crystals weighted up to ~ 16 mg were arranged on the heat capacity puck. One piece of cubic crystal was polished into plate-like shape with the size of 0.64×0.5×0.2 mm$^3$ and was used for electrical resistivity measurements.

Non-magnetic high-pressure cell with 500 μm culet was adopted to apply pressures up to 50 GPa. The tungsten foil was used as gasket, which was preindented to the thickness of 40 μm, and the electrodes are made of 5 μm thick Pt foil. A hole with 200 μm was drilled to hold sample and pressure transmission medium NaCl. The mixture of epoxy and cubic BN was used to insulate sample and electrodes from metallic gasket. The electrical resistivity as well as Hall resistivity measurements were carried out with Van der Pauw method. Electrical resistivity data down to 1.3 K were collected on self-design low temperature cryostat. Electrical transport measurements under magnetic fields were performed on PPMS-9 with temperature down to 1.8 K and magnetic field up to 9 T.

Raman spectroscopy studies were carried out at room temperature. The spectra were collected on a customary confocal micro-Raman spectrometer with unpolarized He-Ne laser (632.8 nm wavelength) as the excitation source. The resolution of single-grating spectrograph is 1 cm$^{-1}$.

**Results and discussions**

The chemical composition analysis of the obtained crystals reveals that the actual atomic ratio of Ir:Bi:Te is 1.04:0.78:1.18 (Table 1). SXRD analysis indicates that the obtained crystals are belong to pyrite-type structure with space group Pa-3 (No. 205) and the refined lattice constant is $a = 6.5004(3)$ Å (Table 2), which is consistent with the literature data [17]. The refinement of atomic occupancy verifies that Bi is partially substituted by Te getting the formula IrBi$_{0.79}$Te$_{1.21}$ (Table 2) in good agreement with the EDX results.

The electron doping from the substitution of Bi by Te with an actual composition IrBi$_{0.8}$Te$_{1.2}$ leads to the metallic behavior of our samples upon cooling (Fig. 1a) in contrast to a semiconducting behavior of reported IrBiTe [18]. Furthermore, the electrical resistivity of IrBi$_{0.8}$Te$_{1.2}$ demonstrates superconductivity below 1.7 K (Fig. 1a). Further evidence for the bulk superconductivity of IrBi$_{0.8}$Te$_{1.2}$ is obtained from the specific-heat jump at $T_c$, and the transition temperature shifts to lower temperature with applying magnetic fields (Fig. 1b). In the normal state above $T_c$, electrons and phonons do contribution to the heat capacity, which can be described by $C_p = \gamma T + \beta T^3$. Based on linear fit of the $C_p/T$ vs $T^2$ curve above $T_c$ (Fig. 1b), the deduced Sommerfeld coefficient $\gamma$ is 4.85 mJ mol$^{-1}$ K$^{-2}$ fu$^{-1}$ (fu represents formula unit), and the calculated Debye temperature $\theta_D$ is 252 K according to $\theta_D = ((12/5)NR\pi^4/\beta)^{1/3}$, at which $N$ is the number of atoms in chemical formula, and $R$ is the molar gas constant. We note that the superconducting transition is broad, and the electron heat capacity jump $\Delta C_e/\gamma T_c$ at $T_c$ is 0.57 which is lower than the BCS weak coupling limit value 1.43. This can be understood by the inhomogeneity in the crystals due to chemical doping (table 1).

Application of external pressure enhances the superconductivity in IrBi$_{0.8}$Te$_{1.2}$ (Fig. 2a). The highest $T_c$ of 2.6 K is achieved at the pressure of 26.5 GPa. With further compression above 26.5 GPa, $T_c$ turns to decrease slightly demonstrating a dome-shaped dependence on pressure (Fig. 2b). In the normal state, the conductivity of IrBi$_{0.8}$Te$_{1.2}$ displays metallic behavior up to the highest applied pressure, although the room temperature resistivity exhibits nonmonotonic behavior at high pressure evidenced by starting to increase as pressure is above ~ 8 GPa (the inset of Fig. 2a).

The resistivity of IrBi$_{0.8}$Te$_{1.2}$ at ambient pressure and at 26.5 GPa was further studied under magnetic fields (Fig. 3a). Magnetoresistance measurements were performed at selected stable temperatures at ambient pressure, and for high pressure data, temperature dependence of resistivity at several static magnetic fields were collected. The superconducting transition is suppressed by applying external magnetic fields (the inset of Fig. 3a). According to Ginzburg-Landau equation $H_{c2}(T) = H_{c2}(0)(1 - t^2)/(1+ t^2)$ (where $t = T/T_c$), the estimated upper critical fields at zero temperature $\mu_0 H_{c2}(0)$ are 1.8 T and 3.9 T for ambient pressure and 26.5 GPa, respectively. Thus, the ratio $\mu_0 H_{c2}(0, 26.5\text{GPa})/ \mu_0 H_{c2}(0, 0 \text{ GPa}) \approx 2.2$ is close to the ratio $T_c^2(26.5 \text{ GPa})/ T_c^2(0 \text{ GPa}) \approx 2.3$ implying that $\mu_0 H_{c2}(0) \propto T_c^2$ as can be expected for single band BCS superconductor [19]. The corresponding Ginzburg-Landau coherence length $\xi(0)$ is about 13 nm and 8 nm calculated from $\mu_0 H_{c2}(0) = \Phi_0/(2\pi\xi^2(0))$ with the flux quantum $\Phi_0 = 2.07 \times 10^{-15}$ Wb.

The electronic properties of the normal state have been further studied by the Hall effect measurements at 10 K under external pressures. Hall resistivity is linearly dependent on the magnetic field with negative Hall coefficient (Fig. 3b), indicating that electrons contribute to the conductivity. This is consistent with the electron doping induced by replacement of Bi by Te in the IrBi$_{0.8}$Te$_{1.2}$ compound. The deduced (based on one-band model) electron density and mobility data are shown in Fig. 3c as a function of pressure. The carrier density monotonically increases upon compression, although the slope varies obviously as pressure increases above ~ 5 GPa. Correspondingly, the mobility increasing at low pressures turns to decrease continuously, getting the maximum value at ~ 5 GPa. The decrease of electron mobility might explain the observed increase of normal state resistivity at high pressure (Fig. 2a).

Pressure-dependent Raman spectroscopy studies have been performed to check the correlation between the superconductivity enhancement and anion bond instability. Raman spectra at selected pressures are shown in Fig. 4a. According to group analysis, five phonon modes are Raman active in pyrite-type compounds, as $A_g+E_g+3T_g$. $A_g$ is attributed to the in-phase and out-of-phase stretching vibrations of X-X dimer. $E_g$ librational mode is attributed to vibrations perpendicular to the X-X bond. $T_g$ involves various librational and stretching motions. The observed Raman peaks are highly broadened comparing with the spectra of binary pyrite-type compounds [20-22]. This can be explained by the random occupancy of Bi and Te, which is also reported in NiS$_x$Se$_{1-x}$ [20,21]. The observed spectra are dominated by extremely broad peaks between 150 cm$^{-1}$ and 200 cm$^{-1}$ (Fig. 4a). Comparing the spectrum of IrBi$_{0.8}$Te$_{1.2}$ with that of other pyrite-type compounds [21,22], the observed broad peak might be attributed to the anion stretching A$_g$ mode of Te-Te, Te-Bi and Bi-Bi dumbbells due to random occupancy of Bi and Te. The profiles of spectra keep the same as pressure increases to 43.3 GPa, indicating that no structural transition occurs under pressure. Moreover, the Raman peak exhibits continuous blue shift as pressure increases. This is consistent with the phonon hardening due to lattice compression (Fig. 4b), as observed in FeS$_2$ [23]. Contrary to PdSe$_2$ and NiSe$_2$ under pressure [4,11], no phonon softening is observed with applying pressure in IrBi$_{0.8}$Te$_{1.2}$. Although the observed Raman frequency monotonously increases with the increase of pressure, a discontinuity in the pressure dependence of frequency is detected at pressure ~ 6-7 GPa (Fig. 4b). This might be associated with the pressure-induced modification of the electronic structure as reported in pressure studies on NiTe$_2$ [7]. Indeed, the pressure is close to that at which the peculiarities of electronic properties are observed (Fig 3c). Thus, the discontinuity in pressure dependence of the Raman frequency might indicate the modification of electron-phonon interaction. However, pressure dependent structural studies are necessary to elucidate the observed peculiar behavior of phonon frequency.

With regard to the effect of pressure on the superconductivity in IrBi$_{0.8}$Te$_{1.2}$, the critical temperature $T_c$ shows, similarly to the superconductivity in pressure-induced pyrite-phases of PdSe$_2$ and PdS$_2$ [4,11], a dome-shaped dependence on pressure with the highest $T_c$ of ~ 2.6 K at ~ 27 GPa (Fig. 2).. Comparing with band-insulator FeS$_2$, extra electrons induced by electron doping partially fill into e$_g$ band and enhance the carrier concentration. Since e$_g$ band hybridizes with *X p* orbital, electron-doping may lead to the instability of the *X-X* anion dimer [24,25]. Both can play important roles in the emerging superconductivity. However, the frequency of phonon in IrBi$_{0.8}$Te$_{1.2}$ exhibits monotonically pressure dependence without structural transition and phonon softening. This suggests that the enhancement of superconductivity in IrBi$_{0.8}$Te$_{1.2}$ shows no correlation with the instability of anion dimer. Thus, it is most likely due to the increase of carrier concentration induced by external pressure. Such pressure-induced superconductivity due to enhanced carrier density is also observed in PtBi$_2$ [9]. The decrease of $T_c$ at high pressure may be due to phonon stiffening. Our investigations elucidate that electron doping induces superconductivity in semiconducting IrBiTe, and charge carrier doping is a practical way to further tune the superconductivity except for applying pressure.

**Conclusions**

In summary, we synthesized ternary superconducting compound IrBi$_{0.8}$Te$_{1.2}$. The self-electron doping with substitution of Bi by Te leads to metallic conductivity in contrast to semiconducting IrBiTe. Furthermore, the superconductivity with $T_c$ = 1.7 K is observed. The effects of pressure on the superconductivity, electronic properties and lattice dynamic of IrBi$_{0.8}$Te$_{1.2}$ were studied. Application of pressure enhances the superconductivity in IrBi$_{0.8}$Te$_{1.2}$, and the $T_c$ exhibits a dome-shaped pressure dependence with a maximum value of ~ 2.6 K at 26.5 GPa. The increase of carrier density by applying pressure plays crucial roles in the enhancement of superconductivity. Our studies on IrBi$_{0.8}$Te$_{1.2}$ suggest that cobaltite family of ternary compounds with pyrite structure might be a good platform for superconductivity to emerge by modifying the chemical composition or applying pressure.

**Acknowledgements**

This work was financially supported by the ERC Advanced Grant no. 742068, "TOPMAT".

**Figure captions**

**Figure 1.** The electrical resistivity and heat capacity measurements at ambient pressure. (a) The metallic conducting behavior evidenced by the decrease of resistivity by cooling, and the inset displays the superconducting transition at low temperature. (b) The heat capacity shown as $C_p/T$-$T^2$ curve exhibits a jump due to superconducting transition. The red solid line represents the linear fit of the normal state data, which gives the Sommerfeld coefficient of 4.85 mJ mol$^{-1}$ K$^{-2}$ fu$^{-1}$, and Debye temperature of 252 K. The inset shows that both the magnitude of heat capacity jump and the transition temperature are suppressed with applying pressure.

**Figure 2.** (a) The temperature dependence of electrical resistivity down to 1.3 K under selected pressures. The upper left inset is the expanded view of normalized resistivity around superconducting transition, and the lower right inset displays the pressure dependence of room temperature resistivity with pressures up to 45 GPa. (b) The pressure dependence of onset $T_c$ obtained from two pieces of crystals. The triangle symbol represents the $T_c$ at ambient pressure.

**Figure 3.** Studies under magnetic fields. (a) The temperature dependence of upper critical field at ambient pressure and the pressure of 26.5 GPa. The solid lines are the fit according to Ginzburg-Landau function. The inset is the typical data displaying that magnetic field suppresses the superconductivity. (b) The typical Hall resistivity linearly fitted with single band model. (c) The pressure dependence of the obtained carrier concentration and mobility from Hall effect measurements.

**Figure 4.** (a) Raman spectral under selected pressures, and the blue curve at bottom is the spectrum when pressure is released to 0.1 GPa. (b) The derived Raman shift of all measured Raman spectra with respect to pressure. The blue square denotes the Raman shift with pressure released, and the red line is guide for the eyes.

**Table 1.** Chemical component of the obtained crystals characterized with EDS.

**Table 2.** The refined lattice parameters of obtained crystals. Space group: Pa-3 (No. 205), lattice constant: $a$ = 6.5004(3) Å. $R_w$: 0.0424, final goodness of refinement: 1.379, and R: 0.0230.

Figure 1

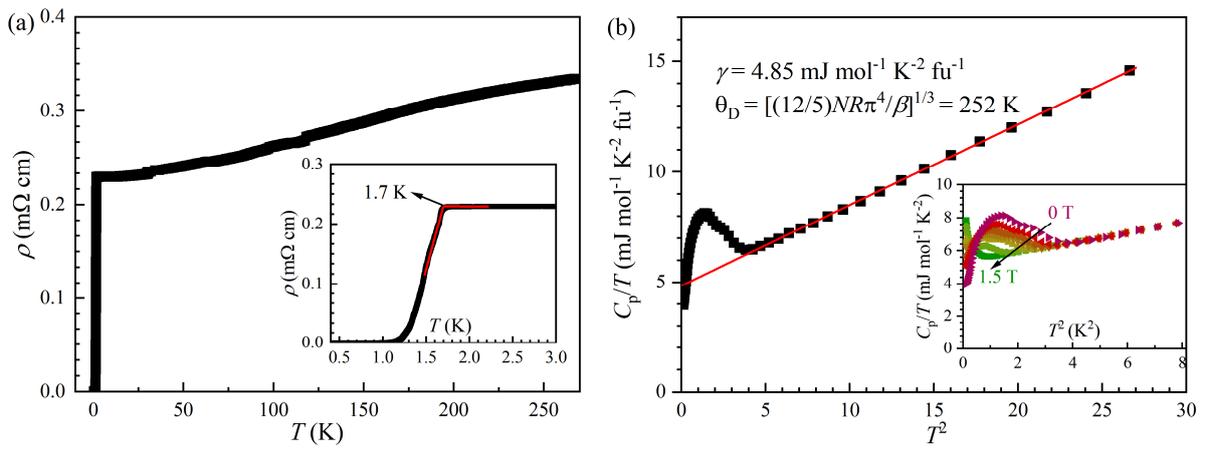

Figure 2

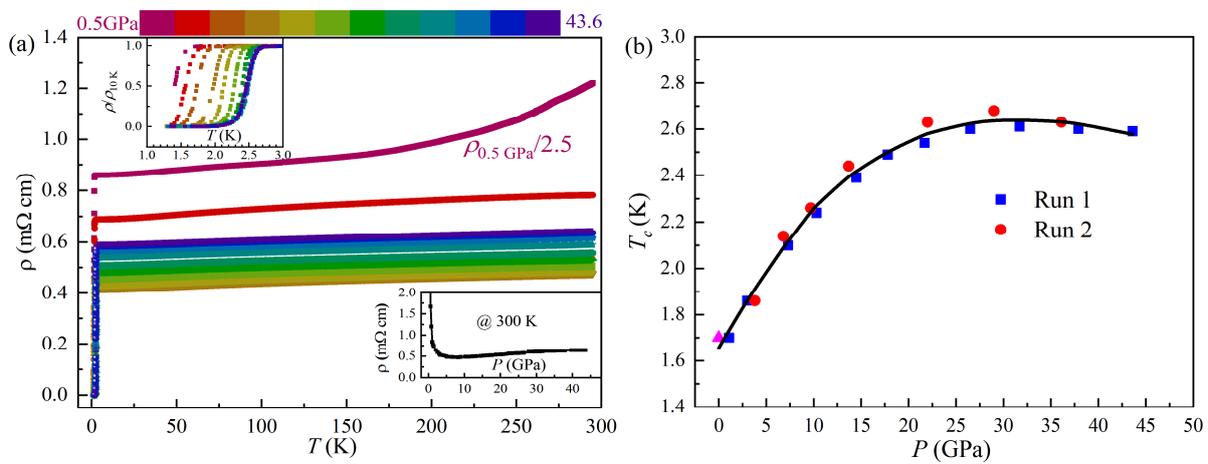

Figure 3

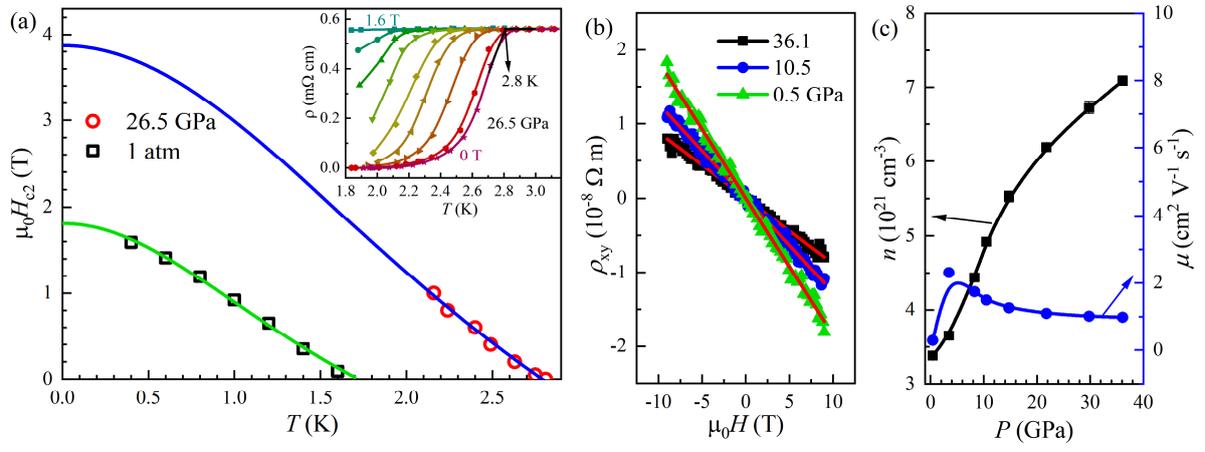

Figure 4

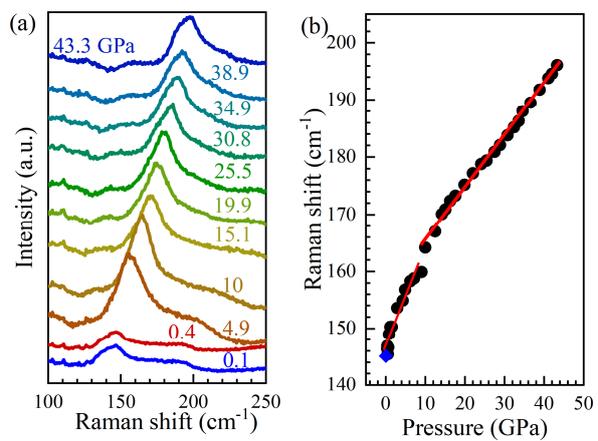

Table 1

| Experiment number | Ir | Bi | Te |
|---|---|---|---|
| 1 | 1.07 | 0.83 | 1.10 |
| 2 | 1.07 | 0.77 | 1.16 |
| 3 | 1.04 | 0.75 | 1.21 |
| 4 | 1.01 | 0.78 | 1.21 |
| 5 | 1.02 | 0.77 | 1.21 |
| 6 | 1.11 | 0.85 | 1.03 |
| 7 | 0.97 | 0.7 | 1.33 |
| Mean value (standard deviation) | 1.04(0.05) | 0.78(0.05) | 1.18(0.09) |

Table 2

| Atom | x | y | z | Occ. | U | site |
|---|---|---|---|---|---|---|
| Ir | 0.5 | 0.5 | 0.5 | 1.000 | 0.00614(10) | 4b |
| Bi | 0.86972(2) | 0.63028(2) | 0.36972(2) | 0.395(6) | 0.00583(10) | 8c |
| Te | 0.86972(2) | 0.63028(2) | 0.36972(2) | 0.605(6) | 0.00583(10) | 8c |